\documentclass[twocolumn]{aastex61}

\usepackage{newtxtext,newtxmath}
\usepackage{multirow}
\usepackage{amsmath}
\usepackage{footnote}
\usepackage{float}

\received{\today}
\revised{\today}
\accepted{\today}
\submitjournal{ApJL}

\shorttitle{Thermohaline mixing in EMP stars}
\shortauthors{Henkel et al.}

\begin{document}

\title{Thermohaline mixing in extremely metal-poor stars}

\correspondingauthor{Kate Henkel}
\email{kate.henkel@monash.edu}

\author{Kate Henkel}
\affiliation{Monash Centre for Astrophysics, School of Physics \& Astronomy, Monash University, Clayton 3800, Victoria, Australia}

\author{Amanda I. Karakas}
\affiliation{Monash Centre for Astrophysics, School of Physics \& Astronomy, Monash University, Clayton 3800, Victoria, Australia}

\author{Andrew R. Casey}
\affiliation{Monash Centre for Astrophysics, School of Physics \& Astronomy, Monash University, Clayton 3800, Victoria, Australia}
\affiliation{Faculty of Information Technology, Monash University, Clayton 3800, Victoria, Australia}

\author{Ross P. Church}
\affiliation{Monash Centre for Astrophysics, School of Physics \& Astronomy, Monash University, Clayton 3800, Victoria, Australia}
\affiliation{Department of Astronomy and Theoretical Physics, Lund Observatory, Box 43, SE-221 00 Lund, Sweden}

\author{John C. Lattanzio}
\affiliation{Monash Centre for Astrophysics, School of Physics \& Astronomy, Monash University, Clayton 3800, Victoria, Australia}



\begin{abstract}

Extremely metal-poor (EMP) stars are an integral piece in the puzzle that is the early Universe, and although anomolous subclasses of EMP stars such as carbon-enhanced metal-poor (CEMP) stars are well-studied, they make up less than half of all EMP stars with [Fe/H] $\sim -3.0$. The amount of carbon depletion occurring on the red giant branch (carbon offset) is used to determine the evolutionary status of EMP stars, and this offset will differ between CEMP and normal EMP stars. The depletion mechanism employed in stellar models (from which carbon offsets are derived) is very important, however the only widely available carbon offsets in the literature are derived from stellar models using a thermohaline mixing mechanism that cannot simultaneously match carbon and lithium abundances to observations for a single diffusion coefficient. Our stellar evolution models utilise a modified thermohaline mixing model that can match carbon and lithium in the metal-poor globular cluster NGC 6397. We compare our models to the bulk of the EMP star sample at [Fe/H] $= -3$ and show that our modified models follow the trend of the observations and deplete less carbon compared to the standard thermohaline mixing theory. We conclude that stellar models that employ the standard thermohaline mixing formalism overestimate carbon offsets and hence CEMP star frequencies, particularly at metallicities where carbon-normal stars dominate the EMP star population.

\end{abstract}



\section{Introduction} \label{sec:intro}

Extremely metal-poor (EMP) stars tell us about the origin and evolution of the Galaxy at the earliest times and consequently the origins of the heavy elements we see today. Studies of EMP stars are particularly useful with regards to chemical evolution and determining the progenitors of early supernovae, and allow astronomers to gain a better understanding of the nature of elusive population III stars \citep{placco14}. Therefore EMP star studies are very useful for understanding both stellar and galactic evolution.

EMP stars can be further categorised based upon their abundances. Carbon enhanced metal-poor (CEMP) stars have an elemental carbon abundance of [C/Fe] $> +0.7$ \citep{aoki07} and comprise 10-20\% of all EMP stars with [Fe/H] $\lesssim -2.0$ \citep{norris13}. \citet{placco14} find this frequency increases to 43\% for stars with [Fe/H] $\leq -3.0$ and 100\% for stars with [Fe/H] $\leq -5.0$. The dominant subclass of CEMP stars are CEMP-s stars (enriched in \textit{s}-process elements), which comprise around 80\% of all CEMP stars \citep{aoki07}. Another category of EMP stars is that of nitrogen enhanced metal-poor (NEMP) stars \citep{pols09}.

On the Red Giant Branch (RGB), low-mass stars (LMS, $\lesssim 2.5$ M$_{\odot}$) exhibit signs of mixing beyond the inner boundaries of their convective envelopes. This produces a decline in the surface abundance of carbon ([C/Fe]) and $^{12}$C/$^{13}$C \citep{briley90,gilroy91,gratton00,gsmith03,martell08} and lithium \citep{lind09}, and an increase in the nitrogen abundance \citep{gratton00}. The effects of extra mixing are observed in LMS in open and globular clusters, though the decline of surface abundances is steeper at lower metallicity \citep{gratton00,aoki07}.

There are a number of postulated explanations for extra mixing, and thermohaline mixing \citep{ulrich72,kippenhahn80} is one popular theory \citep[for others see][]{denissenkov09,denissenkov12,lagarde12b,karakas14dawes}. Thermohaline mixing is driven by the molecular weight inversion created by the reaction $^3$He($^3$He,2p)$^4$He reaction that occurs just above the hydrogen shell, where temperatures are sufficient for $^3$He destruction but are too low for significant contributions from the p+p or CNO cycle reactions \citep{denissenkov04,eggleton06,charbonnel07a}. The decrease in molecular weight is so small that it can only drive mixing in a region that has been completely homogenised, e.g. by first dredge-up (FDU). The depth of FDU reduces as metallicity decreases, therefore the location of the molecular weight inversion is further out in mass in low-metallicity stars but at similar temperatures to stars of higher metallicity \citep{church14}. However the temperature gradient in low-metallicity stars is not as steep as in stars of higher metallicity, and temperatures are consequently higher in the thermohaline region at low metallicities. This produces a steeper decline in abundances, which matches observational trends \citep[an effect also seen in other theoretical stellar models, e.g.][]{charbonnel07a}.

A problem encountered when invoking thermohaline mixing as the extra mixing mechanism on the RGB is the inability to simultaneously match carbon and lithium abundances for a single diffusion coefficient value \citep{angelou15}. Although observations of both carbon and lithium abundances strongly constrain the extra mixing mechanism occurring on the RGB, due to their different burning temperatures, many previous studies analysed only one of these abundances. \citet{henkel17} overcome this problem for the first time by using a phenomenological modification of thermohaline mixing that invokes faster mixing in the hotter part of the thermohaline region and slower mixing in the cooler part. They match their stellar models to observations of the metal-poor ([Fe/H] $\sim -2.0$) globular cluster NGC 6397 \citep{gratton03}.

The amount of carbon depletion due to extra mixing, also called the ``carbon offset'', increases as a star evolves up the RGB and is therefore a function of evolutionary stage \citep{placco14}. Theoretical offsets, the magnitude of which are dependent upon evolutionary status ($\log$ g), are applied to red giant surface abundances to recover initial (pre-RGB) carbon abundances. Applying offsets to a population of stars allows us to derive the frequency of stars that were born with carbon abundances high enough to be classified as CEMP stars, even though the star's current carbon abundance may not be in the CEMP star range. Offsets at a given $\log$ g are determined according to $\text{[C/Fe]}=\text{[C/Fe]}_{\text{initial}}-\text{offset}$. Observers therefore require theoretical models for information on these offsets, yet there are few low-metallicity, low-mass stellar models that include a thorough study of the effects of extra mixing. Existing low-metallicity stellar model grids include \citet{stancliffe09}, \citet{lagarde12b}, and \citet{placco14}.

The models of \citet{stancliffe09} were computed using the STARS evolution code and include the thermohaline prescription of \citet{ulrich72} and \citet{kippenhahn80}. The equation for the thermohaline diffusion coefficient is given by

\begin{equation} \label{eq:dth}
D_t = C_t K (\phi / \delta) {{-\nabla_{\mu}} \over {(\nabla_{\rm ad} - \nabla)}},
\end{equation}

where all variables have their usual meaning, and the dimensionless parameter, $C_t$, is treated as a free parameter. \citet{stancliffe09} use $C_t = 1000$, following \citet{charbonnel07a}. \citet{placco14} also use the STARS code and include thermohaline mixing according to \citet{stancliffe09}. Although \citet{stancliffe09} acknowledge that surface lithium decreases due to extra mixing, both \citet{stancliffe09} and \citet{placco14} focus their discussions on CEMP stars and consequently perform analyses of carbon and nitrogen abundances only. Additionally, \citet{placco14} shift their theoretical models by $\log$ g $= 0.5$ to make the observed and theoretical locations of extra mixing coincide. \citet{lagarde12b} utilise the Ulrich/Kippenhahn thermohaline mixing implementation described in \citet{charbonnel07a} with $C_t = 1000$. Although \citet{lagarde12b} do not discuss surface abundance changes, \citet{charbonnel07a} analyse their theoretical surface abundances of lithium, carbon, and nitrogen and compare to field stars \citep{gratton00}. Their models match carbon but other elements are not well explained, e.g. lithium, which is depleted too much in stellar models compared to the observations.

Although available, large grids of metal-poor stellar models are often coarse in metallicity and fail to adequately model the bulk of the EMP star sample at relevant metallicities. Consequently, observers must either derive empirical relations to determine the carbon offset for low-metallicity stars or use online resources such as the tool developed by \citet{placco14}. This tool for determining carbon offsets for a given set of stellar parameters is based on stellar evolution models that employ the standard thermohaline mixing mechanism. It has been shown that the standard formalism cannot simultaneously match carbon and lithium abundances at metallicities where ``normal'' EMP stars dominate the population \citep[i.e. stars that show no carbon or nitrogen enhancement with $\text{[Fe/H]} > -3$,][]{norris13,placco14,henkel17}. This will affect the carbon offsets and hence the inferred frequency of CEMP stars in the Galaxy.

We apply the methodology of \citet{henkel17} to a subsample of EMP stars from the Stellar Abundances for Galactic Archeology (SAGA) database \citep{suda08}. The SAGA database includes thousands of EMP stars, many of which have carbon, nitrogen, and lithium abundances.


\section{Theoretical models} \label{sec:models}

We compute our stellar models using MONSTAR, the Monash version of the Mt. Stromlo stellar evolution code, and refer to \citet{henkel17} for a detailed description of the code and input physics. Stellar models are evolved from before the zero-age main sequence (Hayashi track) to the helium flash.

We select carbon-normal EMP dwarfs and giants from the SAGA database that have both carbon and lithium observations and find that they have an average metallicity of [Fe/H] $\sim -3$ (metallicity range of sample is $-3.5$ to $-2.5$). Therefore to allow meaningful comparisons with these observations we adopt [Fe/H] $= -3$. We use \citet{asplund09} scaled solar abundances (with the exception of initial $\text{[C/Fe]}$) and an alpha-element enhancement [$\alpha$/Fe] $=0.4$ according to the [$\alpha$/Fe]-[Fe/H] observational trend shown in Fig. 1 of \citet{yong16}.

We construct stellar models with mass $0.8 \text{M}_{\odot}$ and $Y = 0.24$, which produces main sequence turn-off (MSTO) and RGB tip ages of 13.2 and 14.1 Gyrs respectively. Following \citet{henkel17} we include convective overshoot according to \citet{herwig97} and adopt an overshoot factor $f_{\rm OS} = 0.14$ at the formal border of convective regions (defined by the Schwarzschild criterion).

To adequately cover the spread of carbon abundances in the stellar subsample, we test four initial [C/Fe] abundances: [C/Fe] $ = -0.5$, $0$, $+0.3$, and $+0.5$. For each initial carbon abundance, we compare our modified thermohaline mixing scheme that includes a phenomenological temperature dependence \citep{henkel17} to a standard case using the unmodified thermohaline diffusion coefficient equation as derived by \citet{ulrich72} and \citet{kippenhahn80} with $C_t = 1000$. By implementing the modification to the standard formalism detailed in \citet{henkel17}, we facilitate the Cameron-Fowler mechanism \citep{cameron71} for producing $^7$Li in stars by bringing $^7$Be from hotter to cooler regions. This is why our modified models deplete less lithium than the standard thermohaline models but still deplete carbon as required by the observations.

\begin{figure*}
   \centering
   \includegraphics[width=0.9\textwidth]{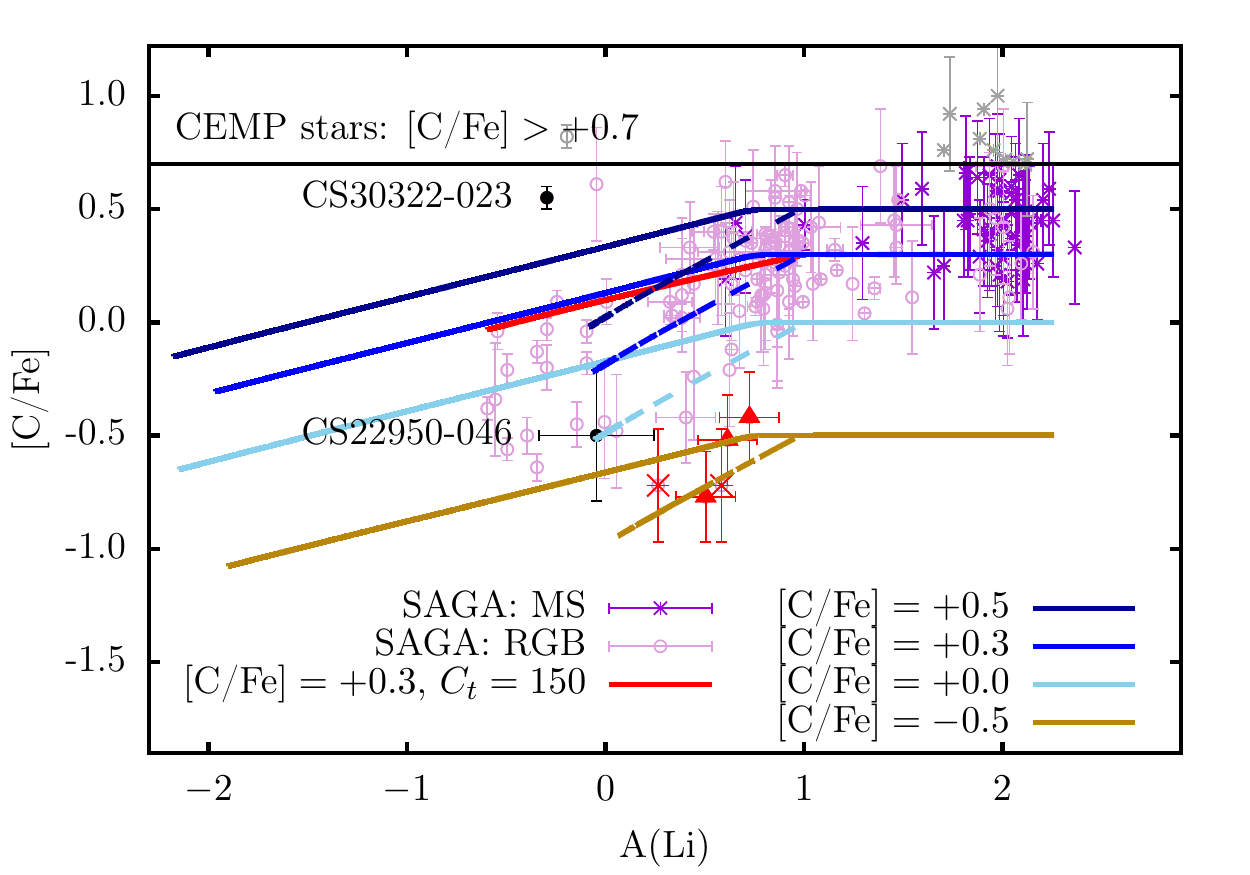}
   \caption{[C/Fe] as a function of A(Li) for our standard and modified thermohaline mixing models with four initial carbon abundances. Observations are from the SAGA database \citep{suda08} with the exception of CS22950-046 \citep{johnson07} and CS30322-023 \citep[the strength of the CN band is due to an overabundance of nitrogen according to][and should therefore be classified as a NEMP star]{masseron06}. The stars in red are discussed further in the text. The criterion for CEMP stars \citep{placco14} is indicated, and stars that satisfy the criterion are shown in grey. For all models, the thermohaline mixing free parameter $C_t = 1000$ except for the red curve where $C_t = 150$ and initial [C/Fe] $=0.3$. Standard thermohaline models are shown with solid lines, and results with the modified algorithm of \citet{henkel17} are shown with dashed lines.}
   \label{fig:licfe}
\end{figure*}

\begin{figure*}
   \centering
   \includegraphics[width=0.9\textwidth]{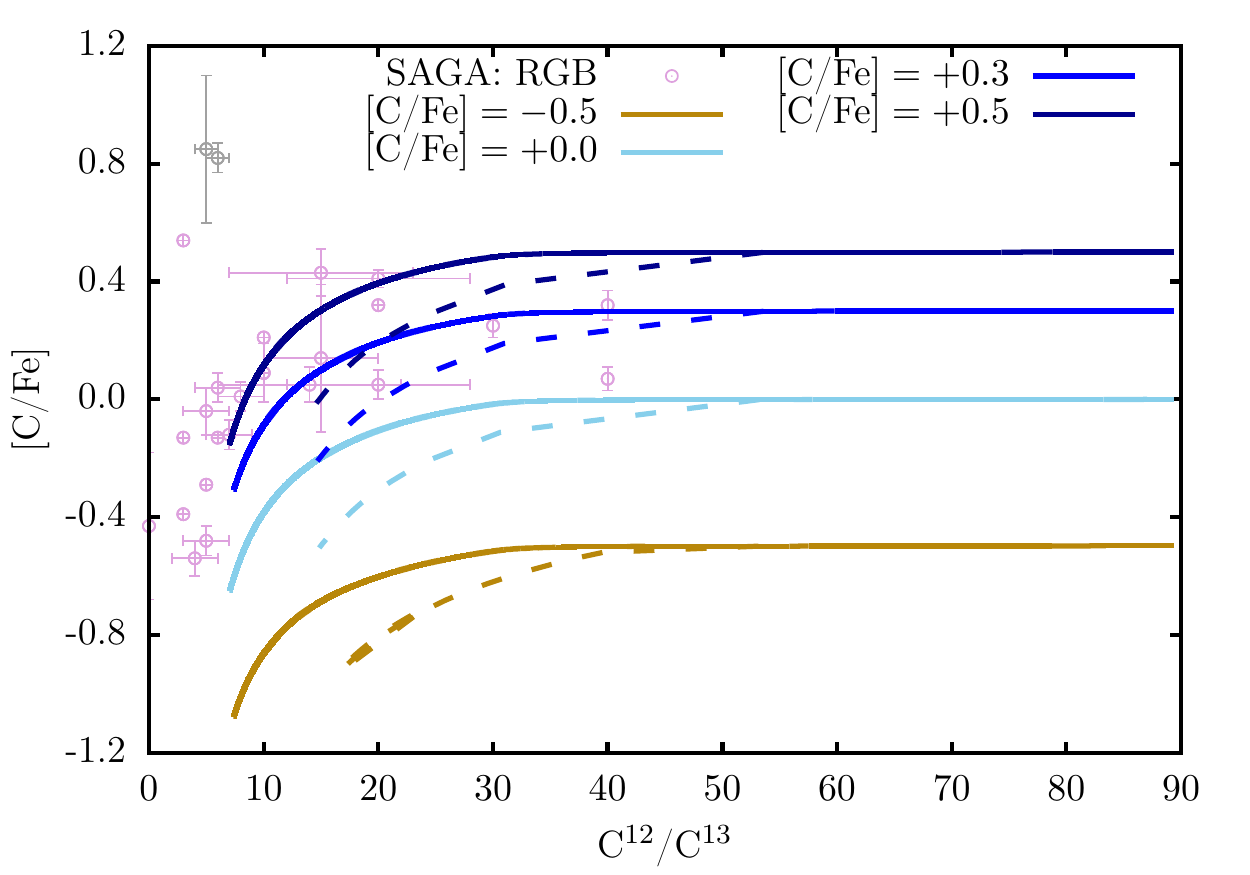}
   \caption{[C/Fe] as a function of $^{12}$C/$^{13}$C for our standard and modified thermohaline mixing models with four initial carbon abundances (details are given in the legend). The stars are those in Fig.~\ref{fig:licfe} that have carbon isotopic observations in the SAGA database \citep{suda08}. The stars that satisfy the CEMP star criterion \citep{placco14} are shown in grey. Standard thermohaline models are shown with solid lines, and results with the modified algorithm of \citet{henkel17} are shown with dashed lines.}
   \label{fig:cfec12c13}
\end{figure*}


\section{Results} \label{sec:results}

The standard thermohaline mixing model cannot reproduce the downturn in [C/Fe] as shown in Fig.~\ref{fig:licfe} (we discuss the stars in red in \S\ref{sec:disc}). Our modified models do not evolve to the low lithium abundances that the standard models do, and do not match the observed stars with A(Li) $< 0$. Stars with abundances as low as the standard models with A(Li) $\lesssim -1$ are not seen in the observed data, however this could be a selection effect or bias towards stars with higher lithium abundances.

Fig.~\ref{fig:cfec12c13} shows that the models with the highest initial carbon abundances are a better match to the observations, and this is expected due to the population being enhanced in $\alpha$-elements. However we limited our selection of stars to those from Fig.~\ref{fig:licfe} with carbon isotopic observations. Due to the limited number of stars with carbon isotopic observations and a possible bias towards stars with higher [C/Fe], it is difficult to confidently conclude which model best matches the data.

\begin{figure*}
   \centering
   \includegraphics[width=0.9\textwidth]{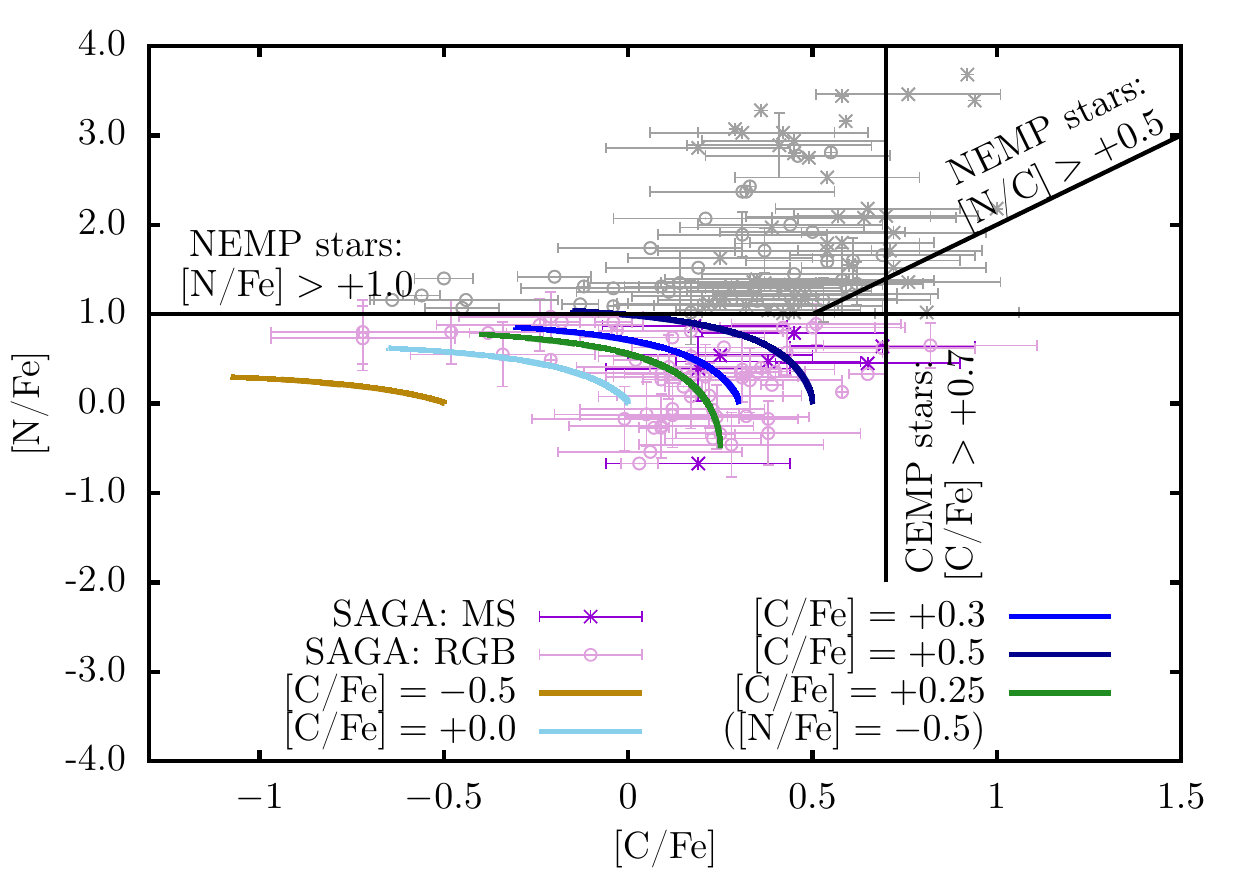}
   \caption{[N/Fe] as a function of [C/Fe] for our standard (solid lines) and modified (dashed lines) thermohaline mixing models with four initial carbon abundances (details are given in the legend). Initial [N/Fe] $=0$ for all models except where indicated in the legend. The standard and modified models overlap because nitrogen has reached saturation. Observations are from the SAGA database \citep{suda08}. The criteria for NEMP \citep{pols09} and CEMP \citep{placco14} stars are shown by the solid lines, and stars that satisfy either criterion are shown in grey.}
   \label{fig:nfecfe}
\end{figure*}

Although the nitrogen abundances of our solution models do not differ from the standard models significantly, as shown in Fig~\ref{fig:nfecfe}, they cover the range of nitrogen observations (when NEMP stars are not considered). The difference between the modified and standard models (for a given initial carbon) is indistinguishable because nitrogen reaches saturation and is unaffected by differences in the mixing mechanism. Indeed, we can achieve a better fit to the nitrogen observations if initial nitrogen is decreased from $0$ to $-0.5$ (shown by the green curve in Fig.~\ref{fig:nfecfe}), but this does not alter our conclusions.



\begin{figure}
   \centering
   \includegraphics[width=\columnwidth]{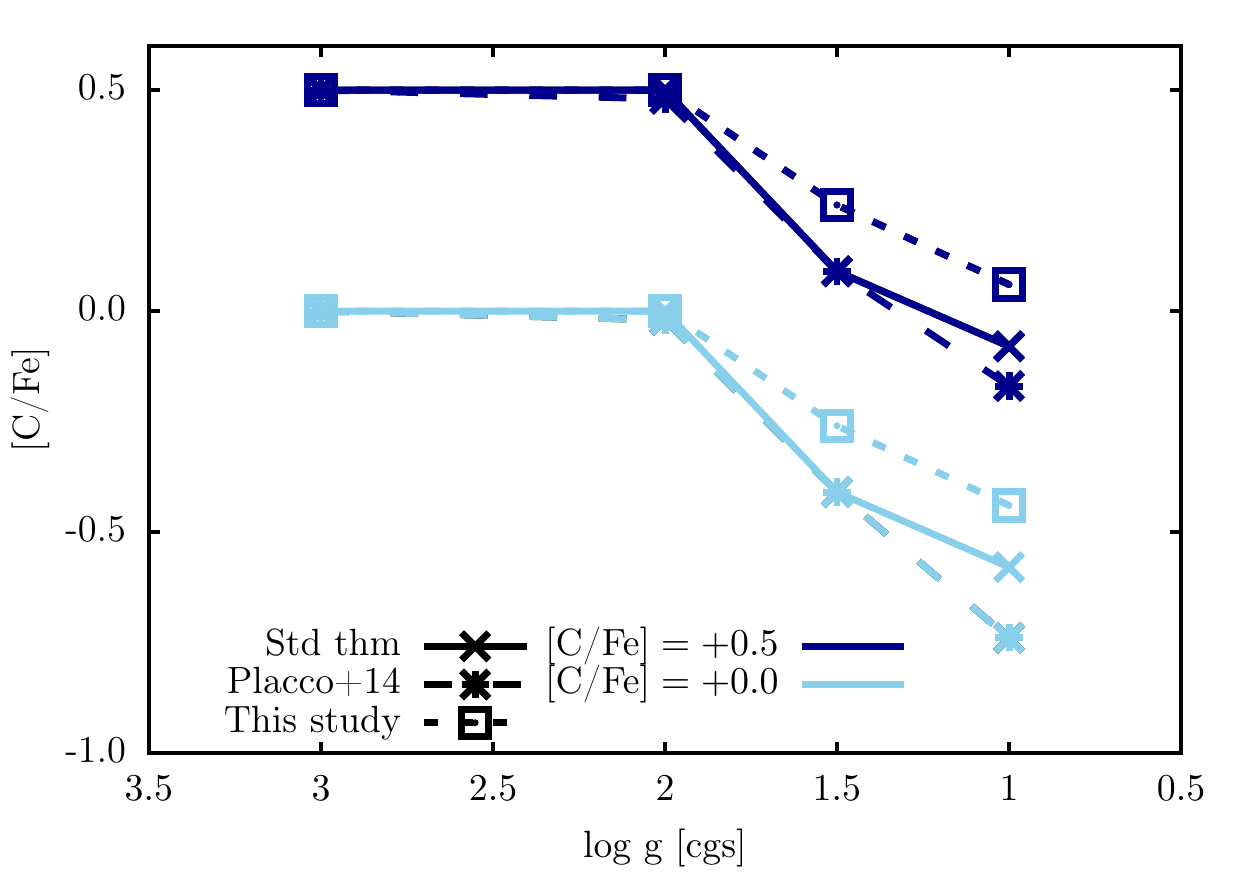}
   \includegraphics[width=\columnwidth]{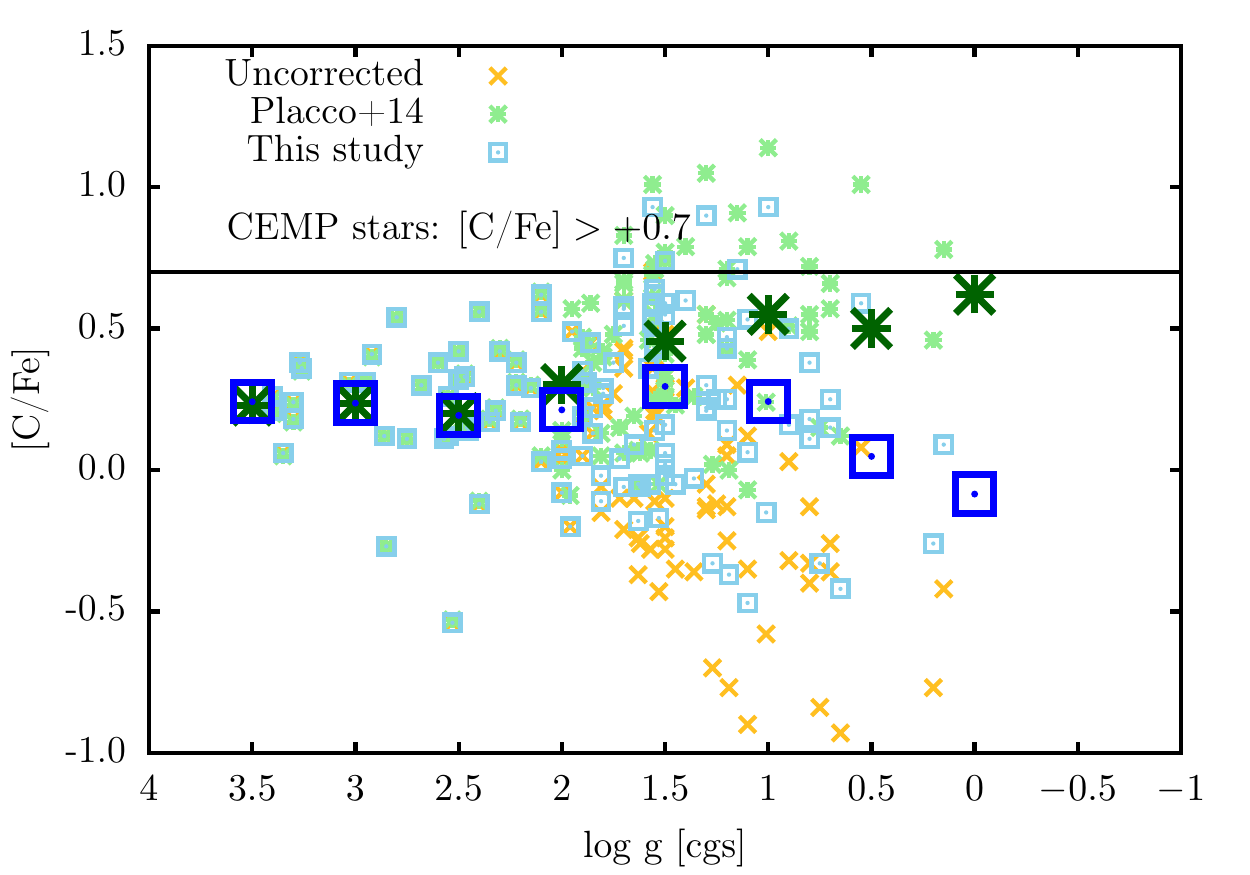}
   \caption{Top panel: The predicted carbon abundance after applying the carbon offsets from the models used in this paper and those of \citep{placco14} for two of our initial carbon abundances (details are given in the legend). Bottom panel: Carbon as a function of $\log$ g of the giant stars in Fig.~\ref{fig:licfe} where the uncorrected carbon abundances are orange crosses, carbon abundances with the offset of \citet{placco14} are green asterisks, and the abundances with the offsets from this study are blue squares. The large symbols are averages in $\log$ g bins (at the $\log$ g value of the symbol $\pm 0.25$) and the symbol colour and shape correspond to the observations in the legend. The CEMP star criterion is indicated \citep{placco14}.}
   \label{fig:carboncorrections}
\end{figure}


\section{Discussion} \label{sec:disc}

Our modified thermohaline mixing models match the observed rate of carbon depletion in metal-poor stars in the [C/Fe]-A(Li) plane (Fig.~\ref{fig:licfe}). This is because our modified models induce mixing that is faster in the hotter region of the thermohaline zone and slower in the cooler part \citep[for more details, see][]{henkel17}.

An issue with implementing thermohaline mixing as the extra mixing mechanism is the requirement for different diffusion coefficient (or $C_t$) values to match carbon and lithium abundances \citep[][]{angelou15,henkel17}. \citet{henkel17} match carbon and lithium abundances for a single value of $C_t$ ($C_t = 1000$) for NGC 6397 giants with [Fe/H] $\sim -2$, however there is no theoretical or observational indication as to whether this is a suitable value of $C_t$ for this sample of EMP stars. Ideally, to determine if a single value of $C_t$ is suitable, we require carbon and lithium abundances as a function of luminosity or absolute magnitude, which are not available for these EMP stars. However Gaia will provide further data (e.g. magnitudes) in the near future. As a test, we determine the surface carbon and lithium abundances of a model that includes the standard thermohaline mixing formalism with $C_t = 150$ and [C/Fe] $=+0.3$ and show the results in Fig.~\ref{fig:licfe}. The model with $C_t = 150$ matches the amount of lithium depletion (unlike our modified models) but only covers the upper envelope of carbon abundances.

\subsection{Carbon-poor/lithium rich giants} \label{subsec:ecpstars}

There are six metal-poor stars that have unusually low carbon for a given lithium abundance, or high lithium abundances for a given carbon abundance. These are represented by the red symbols in Fig.~\ref{fig:licfe}, namely HE1317-0407, HE2253-0849, HE2148-1105A \citep[red filled triangles in Fig.~\ref{fig:licfe}, observed by][]{hollek11}, CS22956-114, and HE1320-1339 \citep[red stars in Fig.~\ref{fig:licfe}, observed by][]{roederer14}. Compared to the dominant trend in the observed stars and our stellar models, these stars appear to have either (1) depleted carbon faster or (2) experienced lithium production.

There are several possible theoretical explanations for these anomolous stars:
\begin{enumerate}
   \item The stars had lower initial carbon, say [C/Fe] $ = -0.5$ (shown by the golden curves in Fig.~\ref{fig:licfe}). No dwarfs have been observed with such a low carbon composition in \textit{this} sample. The lack of observed carbon-poor dwarfs could be explained by selection effects due to the difficulty of measuring low abundances of carbon in dwarfs that have much hotter atmospheres than their cool giant counterparts, and the bias towards observing giants that are much brighter and more numerous than dwarfs.
   \item Thermohaline mixing begins on the RGB and continues during AGB evolution. For this explanation to be explicable with what we observe (i.e. normal lithium and low carbon), we require lithium production by either internal mechanisms such as deep mixing \citep[e.g.][]{stancliffe10,lattanzio15} or external mechanisms such as binary interactions.
\end{enumerate}


After taking into consideration the sources of error that are inherent in deriving abundances from cool evolved stars, we are satisfied that our models match the upper envelope of these anomolous stars. Although beyond the scope of this letter, these stars should be the subject of further investigation.

\subsection{Theoretical carbon offsets} \label{subsec:carbonoffsets}

We now compare the offsets from our models to the theoretical carbon offsets determined by \citet[][detailed in their Fig. 10]{placco14}. The carbon offsets of \citet{placco14} are derived from a grid of 210 models varying in metallicity, initial [C/Fe], and initial [N/Fe], with $C_t = 1000$, and we refer to their paper for a detailed description of their method for determining offsets. Our models in comparison do not vary in metallicity or initial [N/Fe], therefore we only compare to the models of \citet{placco14} that match our initial stellar parameters. We note that \citet{placco14} compute models at [Fe/H] $=-1.3$, $-2.3$, $-3.3$, and $-4.3$, not at our chosen metallicity ([Fe/H] $=-3$). The results for initial [C/Fe] $=+0.0$ and $+0.5$ are shown in the top panel of Fig.~\ref{fig:carboncorrections}, and results for all initial carbon abundances are in Table~\ref{tab:carbonoffsets}.


\begin{table}
   \begin{tabular}{ c c || >{\centering}m{1.3cm} | >{\centering}m{1.3cm} | >{\centering}m{1.3cm} | >{\centering}m{1.3cm} |}
   \hline
   \multicolumn{1}{|c}{\multirow{6}{*}{\rotatebox[origin=c]{90}{[C/Fe]}}} & \multicolumn{1}{|c||}{$0.5$} & \textbf{0.00} 0.00 (0.00) & \textbf{0.00} 0.00 (0.02) & \textbf{0.26} 0.41 (0.39) & \textbf{0.44} 0.58 (0.67) \tabularnewline \cline{2-6}
   \multicolumn{1}{|c}{} & \multicolumn{1}{|c||}{$0.3$}  & \textbf{0.00} 0.00 (0.00) & \textbf{0.00} 0.00 (0.02) & \textbf{0.26} 0.41 (0.41) & \textbf{0.44} 0.58 (0.69) \tabularnewline \cline{2-6}
   \multicolumn{1}{|c}{} & \multicolumn{1}{|c||}{$0.0$}  & \textbf{0.00} 0.00 (0.00) & \textbf{0.00} 0.00 (0.02) & \textbf{0.26} 0.41 (0.41) & \textbf{0.44} 0.58 (0.74) \tabularnewline \cline{2-6}
   \multicolumn{1}{|c}{} & \multicolumn{1}{|c||}{$-0.5$} & \textbf{0.00} 0.00 & \textbf{0.00} 0.00 & \textbf{0.25} 0.40 & \textbf{0.43} 0.57 \tabularnewline \hline
\hline
 & & $3.0$ & $2.0$ & $1.5$ & $1.0$ \tabularnewline\cline{3-6}
 & & \multicolumn{4}{c|}{$\log$ g [cgs]} \tabularnewline \cline{3-6}
    \end{tabular}
  \caption{Carbon abundance offsets in $\log$ g-[C/Fe] space with [Fe/H] $= -3$ and [N/Fe] = 0.0. The bolded values are from our modified models, the unbolded values are from our standard models, and the corresponding theoretical offsets from Fig. 10 of \citet{placco14} are in brackets. \citet{placco14} do not show results for an initial carbon abundance of [C/Fe] $= -0.5$ in their Fig. 10 therefore we do not show their results for this initial [C/Fe] value.}
  \label{tab:carbonoffsets}
\end{table}

Our carbon offsets become larger in magnitude with decreasing $\log$ g (evolution up the giant branch), which is a trend also seen by \citet{placco14} and expected from normal RGB evolution. Additionally, our offsets are typically lower than those of \citet{placco14}. This is because our modified models destroy less carbon on the RGB than models with the standard thermohaline mixing formalism. As our models are a better fit to the observations, we infer that carbon-normal EMP stars at this metallicity destroy less carbon than predicted by the standard theory. Therefore, our offsets should be preferred when inferring the initial carbon abundance of an individual observed star at relevant metallicities.

\subsection{CEMP star frequency} \label{subsec:cempfreq}

Finally, we apply our carbon offsets and the offsets of \citet{placco14} to the data in Fig.~\ref{fig:licfe} and determine the frequency of CEMP stars. To do this, we linearly interpolate in [C/Fe] and $\log$ g, and apply the offset to the observed data to yield ``corrected'' abundances (i.e. inferred birth abundances). We then take the average of the abundances in $\log$ g bins, and in the bottom panel of Fig.~\ref{fig:carboncorrections} we plot the uncorrected and corrected abundances as well as the averages as functions of $\log$ g. The corrected distribution should be relatively flat over all $\log$ g if the correct initial carbon abundances were recovered.

The bottom panel of Fig.~\ref{fig:carboncorrections} shows that the data with our offsets produce a flatter trend than the offsets of \citet{placco14} because our carbon corrections are smaller at low surface gravities. Using the definition of CEMP as stars with [C/Fe] $> +0.7$ \citep{aoki07}, our offsets imply that this sample initially contained $\sim 5.2\%$ CEMP stars, whereas the offsets of \citet{placco14} yield $\sim 14.7\%$. Offsets derived from stellar models that include the standard formalism of thermohaline mixing overestimate the amount the carbon depletion in stars at metallicities not dominated by CEMP stars (i.e. [Fe/H] $\geq -3$), and applying such offsets will consequently yield overestimated CEMP star frequencies at these metallicities. We acknowledge however theoretical and observational CEMP star frequencies are uncertain \citep[for more on this, see][]{cohen05c,frebel06,collet07,pols09,izzard09}.

\section{Conclusions} \label{sec:conclusions}

We have produced stellar models that employ a modification to the standard thermohaline mixing formalism on the RGB and compare our models to a subsample of EMP dwarfs and giants from the SAGA database with [Fe/H] $\sim -3$. Our modified models match observations of both the carbon and lithium abundances of the EMP stars considered.

We determine the amount of carbon depletion (carbon offset) for our theoretical models and find that our carbon offsets are lower than those of \citet{placco14}. This suggests that the offsets suitable for this sample of EMP stars are not as large as the offsets predicted by models that include a standard thermohaline mixing implementation.

We conclude that analysis of EMP stars requires updated carbon offsets, because currently available offsets overestimate the amount of carbon depletion on the RGB and observers using such offsets will overestimate evolutionary burning and mixing. An updated CEMP star frequency has direct implications for Galactic chemical evolution and requires further investigation at varying metallicities, particularly metallicities where CEMP stars are not the dominant class of EMP stars. We also emphasise that a more statistically complete sample of low-metallicity stars is required to gain further insights into this problem.


\bibliographystyle{aasjournal}

\end{document}